# Survey: Mobility Management in 5G

A.N. Kasim, *Istanbul Technical University*

*Abstract*— **For the next wireless communication systems, the major challenges are to provide ubiquitous wireless access abilities, and maintain the quality of service and seamless mobility management for mobile communication devices in heterogeneous networks. Due to rapid growth of mobile users, this demand becomes more challenging where the users always require seamless connectivity while they move to other places at any time. As the number of users increase, the network load also increases, the handover process needs to be performed in an efficient way. However, in many situations, the handover blocking, and unnecessary handover frequently happen, then affect the network and reduces its performance. The problem arises with the movement of mobile user between base stations while the link connectivity becomes weaker and the mobile node tries to switch to another base station to have a better link quality during a call for higher QoS (Quality of Service). In this survey it is aimed to gather together the studies that helps to improve mobility management processes in 5G (Fifth Generation) cellular networks.**

## I. INTRODUCTION

WIRELESS and mobile communication is generally considered as a recent technology, and its basic elements were introduced in late 1980s. From its launching, mobile wireless networks have changed from analog calls to modern architectures. Today, these technologies lead us to utilize high quality wireless network services with high data rates up to hundreds of Mbps. And, it seems not possible to live without networking, mobile devices are now our indispensables. It is expected that the number of the devices connected to network will touch 24 billion at the end of 2020 [1]. These devices are connected through various technologies such as Wi-Fi, LTE (Long Term Evolution), WiMAX (Worldwide Interoperability for Microwave Access), 5G, etc. Yesterday, today, and tomorrow, the wireless networking technologies had to evolve, have to be enhanced, and will have to advance in order to meet with the demands of the human being. Because of that, the cellular network technology has evolved from 1G to 5G in a few decades. The outstanding advancement of wireless networking technology have led a broad emergence of the new high-quality services. At the same time, with the rapid growth in number new mobile devices like tablets and smart phones, network traffic issues have come up. This issue has been a crucial challenge that should be handled since its development. As demand increase, service providers need to give higher quality, lower latency services with higher data rates. Today, this technology again is evolving to a more advanced one which is 5G. Due to the higher capacities, low latencies, dense networks, higher data rates, the traditional wireless network technologies that have been utilized until 4G, have become inadequate and incapable to fulfill the requirements of the 5G cellular network technology. To meet the requirements such as scalability, traffic management, seamless mobility, resource management, etc. novel approaches and methods are needed [2, 3]. Therefore, although 5G wireless technology provides reassuring solutions, it comes with its some challenges like mobility management. Mobility management is basically an operation of GSM or UMTS networks. It is responsible for identifying the locations of the subscribers and following them. Since it is not possible to connect to the network from anywhere without mobile management, it is considered as fundamental element of wireless networks. Thanks to mobility management operations, uninterrupted connection, reliability, security, and high performance were provided.

The rest of the paper is organized as following: The evolution of the wireless communication networks is discussed in section II. In section III, the information about 5G structure is given. In section IV, mobility management, handover operations, and some mobility related studies were examined. Mobility management operations were introduced in section V. And, mobility management in 5G related studies were provided and discussed in section VI. Lastly, the conclusion part was set to section VII.

## II. EVOLUTION OF MOBILE NETWORKS

Wireless mobile network systems that mobile devices utilize has developed over many years. The technologies are presented below:

**1G:** First generation is the earliest type of technology developed for mobile wireless communication. That system that was launched in 1980s provides up to 2.4 kbps data rate. It was a primitive wireless network having only voice call capability. Nevertheless, this technology comes with some drawbacks like poor coverage, low sound quality, low network capacity, low security due to absence of encryption, insignificant handover processes, compatibility problems between systems, etc. The need for more developed system leads 1G to evolve to 2G [4].

**2G:** This technology in which digital systems have been utilized was launched towards to 90s. This digitalization gave rise to mass adoption of mobile communication thanks to SMS and MMS. That can be considered as a kind of revolution in a sociological way. Although it had provided up to 9.6 kbps at the beginning, a few advancements (e.g. GSM, GPRS, EDGE) brought remarkable increases in data rate. The main benefit of the technology is using low power radio signals, which extended the battery life. GSM (Global System for Mobile Communications), GPRS (General Packet Radio Service), CDMA (Code division Multiple Access) are outstanding concepts in this generation. The system in which GPRS, EDGE, and CDMA were implanted is considered as 2.5G [4, 5].

**3G:** 3G mobile communication network scheme was launched in 2000s. In this technology, standardizing the network protocols is also addressed. By standardization, to access data from anywhere as packets became possible. This enhancement led to international roaming services. In early period it was providing data rates up to 2 Mbps. QoS, voice quality, roaming are important advancements of 3G cellular network scheme. On the other hand, power requirements and the consumption are the disadvantageous side of this scheme. Hence, it becomes more expensive to implement and to run the system [4, 5]. CDMA, WCDMA (Wideband CDMA), UMTS (Universal Mobile Communications Systems) are distinguished concepts of 3G. As a result of further improvements, the data rate 5-30 Mbps were provided by the network that is called as 3.5G.

**4G:** 4G was firstly launched in 2009 as LTE (Long Term Evolution) 4G standard. This cellular network scheme is significantly improved version of 3G and 3.5G. Advanced LTE were utilized by 3GPP and the system was standardized with the WiMAX (Worldwide Interoperability for Microwave Access) in order to enhance the mobile communication performance. 4G scheme in which IP (Internet Protocol)-based solutions were integrated, provides high quality voice, data, internet service with high data rates up to 1Gbps for stationary users. This fast mobile access presents HD videos, HQ video conferences, gaming experiences, etc. with low latency and high data rates.

**5G:** In recent years, the demand of data, high data rate, seamless mobility services, and enhanced QoS remarkably increased. These demands brought a reason for development of 5G technology. This mobile communication technology that was started to join our lives in 2020, consists of many novel techniques. Owing to 5G higher data rates, wider coverage, higher capacity, advanced mobility, etc. will be provided. These efficient and high-quality communication services in 5G networks are resulting from also utilizing the available 30-300 GHz frequency bands. The wider available bandwidth enhances the network performance which allows users to use high quality services [6].

## III. 5G STRUCTURE

### A. Design

In 3GPP, an explicit description was given for 5G network structures to enhance cellular communication further. Between the control plane functions, service-based communication models were utilized in the 5G network architecture. Some essential approaches and concepts were adopted in the system. Some important ones can be counted as reduce in dependencies between Core Network (CN) and Access Network (AN), integrating diverse access types, UP (User Plane) and CP (Control Plane) separation which provides flexibility in deployment and scaling, supporting the concurrent access which allows low-latency applications, etc.[7]

### B. Reference Model

The network functions that are adopted in the 5G mobile network architecture will be observed in this part. To begin with, AMF (Access and Mobility Management Function) is a basic component of 5G network. Its tasks consist of registration management, connection management, mobility management and more. Another fundamental element of the network is SMF (Session Management Function) that is responsible for controlling session context, editing PDU (Protocol Data Unit) sessions, and connecting to decoupled data plane. Another core element of the 3GPP 5G system is UPF (User Plane Function) whose task includes connection between DN and infrastructure, providing PDU session anchor point, routing and forwarding, managing QoS, etc. Other primary components and functions of the network can be counted as AUSF (Authentication Server Function), AF (Application Function), DN, NSSF (Network Slice Selection Function), PCF (Policy Control Function), UDM (Unified Data Management), and UE (User Equipment) [7, 8]. The architecture reference model of the 3GPP 5G network is shown in Figure 1 [8].

Fig. 1. 5G Network Architecture

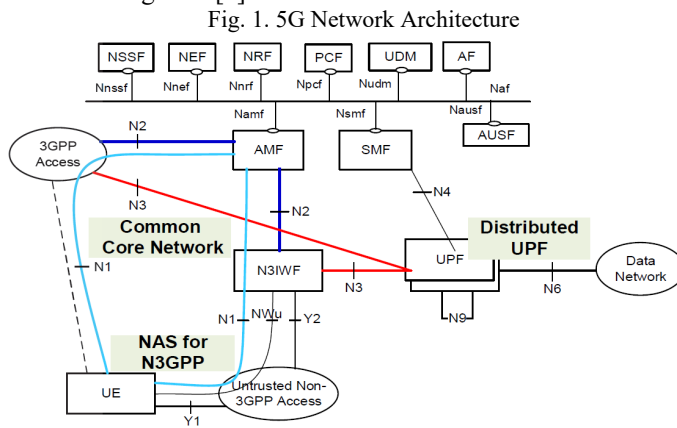

## IV. MOBILITY MANAGEMENT

Basically, mobility management is a function belonging to GSM or UMTS cellular networks. The primary task of the mobile management function is detecting the locations of the users and tracing them in order to provide cellular network service. Actually, mobility is the feature that enables users to connect to the network from anywhere. In a case without mobility management, subscribers would need to change their services or SIM cards when they change their location. That is, owing to mobility management functions, connection and communication constancy, reliability, and performance are provided.

### A. Location Management

Location management function is responsible for location update and paging. The UE notifies the network that it belongs to about its location by transmitting signals. That is, MS (Mobile Station) gives location information to the network according to update procedures. These update signals might be sent only when location is changed or periodically. By evaluating these location signals, the network operates call delivery or paging process.

## B. Handover Management

Handover management is the system guaranteeing that mobile station has continuous connection to the network during mobility. That is, it is a process that manages routing data packets or the connection between MS and network from an AP (Access Point) to another AP. Handover is performed in order to refresh the receiving signal, balance the load, reduce the cost, decrease energy consumption. The whole handover process can be described in three stages. First is initiating handover, second is preparation for handover. In the preparation step, network should make a decision about new target resource for connection and the operations needed to establish new connections. And, the last stage of the handover process is the execution of the operation. This execution should be held according to connection protocols and handover procedures to fulfill the QoS requirements. Since there are plenty of different possible scenarios for handover, the type of the parameter, handover procedure, and the parameters are needed to be defined. Signal interference, power requirements, QoS requirements, efficient allocation of resources, reliability, and robustness must be considered in order to build a proper handover management. The classification of handover processes can be according to the network types, frequencies, techniques, and controller type.

### 1) Handover Types

#### i. Handover Types Based on Networks

**Horizontal Handover:** Horizontal handover is the type of handover that is occurred in homogenous technologies, that is, performed between same access technologies e.g. Wi-Fi to Wi-Fi, GSM to GSM, WiMAX to WiMAX, etc. Horizontal handover is also known as intra-system handover.

**Vertical Handover:** When the handover is occurred in heterogenous technologies, that is, operated between different access technologies e.g. Wi-Fi to GSM, GSM to WiMAX, LTE to Wi-Fi, etc. Vertical handover is also known as inter-system handover.

#### ii. Handover Types Based on Frequency

**Intra-frequency Handover:** It is the type of handover performed between two different access points operating with same frequency bands.

**Inter-frequency Handover:** The handover process occurred between two different access points with different frequency bands is called as inter-frequency handover.

#### iii. Handover Types Based on the Techniques

**Hard Handover:** It is a type of handover method in which all the wireless connections with UE are detached before the new ones are built. That is, in this handover type, process is held as break-before-make which causes the loss of continuity of the communication link. The hard handover may be seamless or not.

**Soft Handover:** Soft handover is a handover technique that is performed according to make-before-break principle. That is, the new connections between UE and the network were set up before breaking the old connections. Hence, both connections run at the same time for a while.

#### iv. Handover Types Based on the Controller

This classification is used when the second and third stages of handover, planning and execution stages, are operated by mobile station or a controller. There are three types of handover such as NCHO (Network-controlled Handover), MCHO (Mobile-controlled Handover), and Mobile-assisted Handover. In NCHO, while handover initiation stage is performed by mobile station, the decision step is held by a controller or the network. That is to say, firstly, the UE initiates the handover process, determines the target AP and inform the network that decides whether the handover is performed or not. In MCHO, the both initiation and decision stages are controlled by the mobile node. On the other hand, in a MAHO process, mobile nodes collect all of the data like SINR (Signal to Interference-plus-Noise Ratio), RSSI (Received Signal Strength Indication), error rate, etc. and send them to the network that chooses the best target points [9].

### 2) Handover Parameters

There are many parameters used in handover process to improve the system. Some parameters are taken from the system to decide about obligation of handover process, some parameters are set up in order to control the process, and some are utilized as performance measurement metrics (HPIs, Handover Performance Indicators.). Hence, they can be investigated under two categories; system parameters and control parameters. RSRP (Reference Signal Received Power) and SINR fall under system parameters group. And, control parameters group includes Hysteresis (RSS Handover Margin), distance handover margin, and TTT (Time-to-Trigger). The parameters like handover failure ratio ($HPI_{HOF}$), ping-pong handover ratio ($HPI_{HPP}$), and call dropping ratio ($HPI_{DC}$) are involved in performance indicators class.

### 3) Related Works

Most of established handover systems are SINR and RSRP based [10]. It is studied in [11], handover between Wi-Fi and WiMAX networks, the handover to Wi-Fi is more preferable in case Wi-Fi exists, because Wi-Fi provides higher bandwidth at lower costs. Whenever UE starts to lose Wi-Fi signal, the handover process probability increases. These decision algorithms are conducted according to SINR, RSRP/RSSI metrics. Also, in [12], researchers propose a proper handover system by utilizing SINR, bandwidth, and RSSI measurements. In the article, it is presented that SINR based vertical handover operation gives optimum result to the user. Nevertheless, performing vertical handover from one operating lower frequency to another operating higher frequency might not be advantageous to the network. In these cases, SINR fluctuations and power consumptions need to be considered. Otherwise, the unnecessary handover operations will be significant drawbacks for the network.

## V. Mobility Management in 5G Networks

Throughout last decades, cellular networks have become disorderly spread on the globe. To organize the wireless networks, mobility management (e.g. handover management) is utilized. To mention about mobility management in LTE networks, this type of network utilizes only hard handover. As mentioned before, hard handover process run in break-before-make principle and it causes some significant issues in mobility management processes. In order to provide a continuous connection to UE, eNB need to support as LTE does not include an RNC (Radio Network Controller) entity [13]. On the other hand, because of massive increase in data traffic, grown in demand the LTE structures will not be applicable for future network scenarios. These current methods will be inadequate for cases of 5G future networks [14].

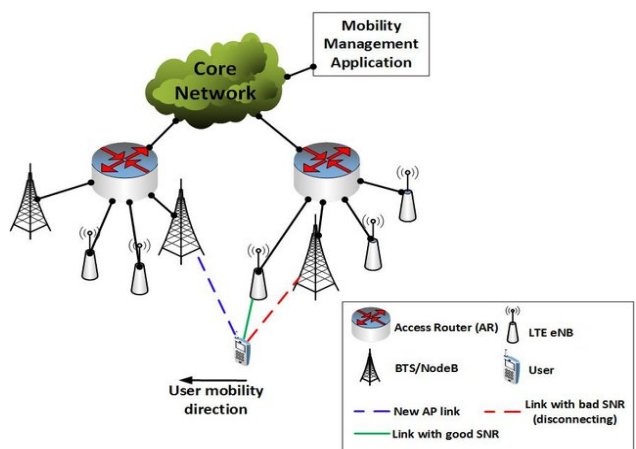

Fig. 2. Mobility Management in 5G

As the use of 5G networks is more adopted, the main differences between 4G and 5G networks will be the outstanding benefits owing to mm-wave frequency bands, beam directional antennas, higher data rates, wider coverage, lower costs, higher capacities, etc. The mobility management scheme for 5G networks shown in Figure 2 [15]. The mobility management services in 5G can be provided on cloud systems. 5G technology that is a packet switched system with outstanding results provides a more efficient and higher performance communication opportunities. And, users are able to utilize the technology and broadband internet connection by their mobile phones [16].

## VI. 5G Mobility Management Related Studies

In this part of the article, the mobility management and handover process related researches are stated and discussed.

Based on the benefits of the handover process, in [17], Qiu *et. al. proposed* a unique innovative method to implement the virtualized network functions and fog computing cases. The research indicates the benefits of virtualizing the network functions. These benefits play an important role to enhance the flexibility and the robustness of the network. The study was carried out by utilizing fog-computing based APs and X2-based handover design.

[18] that is written by Yang *et. al.*, addresses the usage of wireless communication networking for the case of everyday situations. A handover operation for specific cases is presented in the study. The cases about that UE has weak connections with eNBs because of the surrounding buildings and the distances. These obstacles are handled by forwarding the signals from a relay station. By this way, coverage area and communication range increases, which eliminates the interference signal.

Arshad *et. al.,* discusses about enhancement of spectral efficiency and resource allocation operations in a case involved multiple subscribers, in [19]. It is stated that, spectral efficiency can be improved by considering the BS footprints. The handover operation is utilized to increase the 5G network performance. Also, it is stated that handover rate is a significant element that has notable impacts on the network performance and it is needed to be considered appropriately. Moreover, the difficulties suffered while operating handover process in 5G networks are stated and discussed explicitly. A method that minimizes the unnecessary handover rate is proposed in the article, which is stated as topology aware handover approach. This proposed method is verified for single and two-tier networks for downlink connections.

In [20], Barua *et. al.,* proposes a new way to yield better networking performance by utilizing D2D communication method. The proposed D2D method does not need any BS for communication between the UEs. Since mobility management processes are real challenging tasks in D2D communications, a few methods are stated to handle these tasks. The new approach proposes to utilize Time Division Duplex in LTE-A systems. By this way, it becomes possible to implement well-known power control algorithms for Time Division Duplex. However, latency, complexity, and the power issues still go on as important challenges.

Wu *et. al.,* in [21], studies on optimization of the handover parameters in small-cell deployed 5G multi-tier cellular networks. Initially, the article states that present methods used for enhancing QoS are based on former information and the network procedures. And, in case of insufficient information, the mentioned currently used methods are unsuccessful to fulfill the QoS and performance requirements. In purpose of handling this problem, dynamic fuzzy Q-Learning algorithm is proposed as a novel and unique method that ensures continuous D2D communication, lower latency, and lower signaling overhead.

Zhang *et. al.,* in [22], study on QoE (Quality of Experience)-based handover delay reduction on WLAN (Wireless Local Area Networks). In the article, the important effects of WLANs are explicitly stated. In WLANs, subscribers stand generally in the coverage area of numerous networks. And, this complex situation makes handover process between APs more challenging due to handover delays. The handover delays cause QoE to decrease in applications requiring high data rate and low latency. Since the high delays can induce not to meet with performance and QoE requirements, a proactive method for handover management is proposed in the paper. This method utilizes a type of periodical scan-based approach to analyze WLAN information with RSSI. Furthermore, in the study, variable bitrate video coding (VBR) is utilized to enhance the latency performance. Although these methods are discussed for WLAN, it is encouraged to utilize these techniques in 5G cellular networks.

Choi et. al., in [8], stated that, one of the main purposes for architectures is to provide seamless mobility management service by corelating the core network with multiple APs properly. In the paper, approach of MAPDU (Multiple Access Protocol Data Unit) session to manage the data communication in 5G cellular network, and a dynamic mobility management process between various APs are presented. The dynamic anchoring MM method is proposed with End Marker to ensure the connection while UE is moving.

Lastly but not least, in [23], Calabuig *et. al.* states that traditional mobility and resource management methods like increasing spectral efficiency are inapplicable in 5G networks due to the high capacity. They analyze the most promising methods that are defined in METIS project. They utilize context information in their structure to provide a reliable and power efficient mobility management service. Although the proposed solutions seem to enhance mostly resource management, the combination of the methods provide robustness to unplanned cell deployments, mobility for users and cells.

## VII. CONCLUSION

In the last decades, there has been a rapid increase in demand for wireless communication that cellular communication is seen the major part of. And, the latest cases in communication need throughput with the best performance. Because of the innovative differences in 5G technology, it becomes inapplicable to utilize previous traditional mobility management strategies that are used in LTE systems. Therefore, in order to handle the mobility issues in 5G wireless networks, various techniques have been developed. In this study, a comprehensive survey about mobility management in 5G networks is presented after discussing the evolution of wireless networks and mobility management process.